\documentclass[prl,showpacs,twocolumn]{revtex4}
\usepackage{amsmath}
\usepackage{subfigure}
\usepackage[pdftex]{graphicx}
\usepackage{graphics,color}

\begin{document}

\title{Noise-induced dynamical transition in systems with symmetric absorbing states}
\author{D.~I.~Russell}
\author{R.~A.~Blythe}
\affiliation{SUPA, School of Physics and Astronomy, University of Edinburgh, Mayfield Road, Edinburgh, EH9 3JZ, UK}

\date{\today}

\begin{abstract}
We investigate the effect of noise strength on the macroscopic ordering dynamics of systems with symmetric absorbing states. Using an explicit stochastic microscopic model, we present evidence for a phase transition in the coarsening dynamics, from an Ising-like to a voter-like behavior, as the noise strength is increased past a nontrivial critical value. By mapping to a thermal diffusion process, we argue that the transition arises due to locally-absorbing states being entered more readily in the high-noise regime, which in turn prevents surface tension from driving the ordering process.
\end{abstract}


\maketitle

Model systems with \emph{absorbing states} have featured prominently in the quest for a fundamental understanding of far-from-equilibrium phase transitions and critical phenomena \cite{HinBook,OdorBook}.  An absorbing state is one that, once entered, cannot be exited.  An excellent example is the extinction of a species.  Further instances can be found in the dynamics of catalytic reactions \cite{Ziff1986} and of calcium channels in living cells \cite{Baer2000}.

A clear picture of how systems with \emph{multiple} absorbing states behave is still emerging. These include cases where several competing species can drive each other to extinction.  The static phase diagram for systems with two symmetric absorbing states \cite{Dornic2001,Hammal2005,vazquez2008,GallaVM} has recently been shown to have a much richer structure than those with a single absorbing state \cite{HinBook}, featuring in particular order-disorder transitions of several different types \cite{Hammal2005}.

In this Letter, we turn to the \emph{dynamics} of these systems, focusing on how an absorbing state of global order is reached over time.  We report the existence of a dynamical phase transition between two distinct macroscopic ordering regimes at a nontrivial value of the noise amplitude. More precisely, as the scale of local fluctuations in the dynamics is increased (whilst holding fixed all deterministic contributions) the surface tension that normally drives the coarsening process suddenly vanishes, leaving fluctuations as the only means to order. These two ordering modes are characteristic of the Ising and voter models respectively \cite{Hammal2005}. The physical origin of this transition is elucidated by mapping the nonequilibrium stochastic dynamics onto a thermal diffusion process, a technique previously used for models with a single absorbing state \cite{Conrad}. We find that strong noise allows locally-absorbing states to be entered easily. Since the deterministic forces due to surface tension do not act in these states, further order can only be attained through interfacial fluctuations.

This noise-induced dynamical transition is intriguing for several reasons. First, it has not been pre-empted in models with symmetric absorbing states: a renormalization group treatment \cite{GallaVM} suggests that noise is an irrelevant parameter as it is in the equilibrium theory of phase ordering kinetics \cite{BrayRev}. Whilst specific microscopic models  exhibit a wide range of phenomena \cite{HinBook,OdorBook,vazquez2008,GallaVM}, the relative strengths of the deterministic and stochastic contributions to the dynamics are often not independent. Here, we construct a model that avoids this shortcoming, and thus show that a change in noise strength alone can induce a dynamical phase transition. As we discuss in more detail below, our model is applicable to human language behaviour \cite{hudsonkam,Baxter2009} in which noise strength relates to a memory decay rate \cite{Baxter2009}. Thus, the spatial structure of linguistic diversity could in principle be nontrivially affected by how quickly humans forget. Finally, our findings bear on basic questions about nonequilibrium universality classes, a point we return to in the conclusion.

Our starting point is the Langevin equation proposed in Ref.~\cite{Hammal2005} for the general class of systems with two absorbing states. It reads
 \begin{equation}
\partial_{t} \phi = (a\phi-b\phi^{3})(1-\phi^{2})+D\nabla^{2}\phi + \sigma \sqrt{1-\phi^{2}} \hspace{1mm} \eta(t) \;, \label{HammalLE}
\end{equation}
where $\phi \in [-1,1]$ is a continuous coarse-grained field, usually the magnetization. The first term is a deterministic force that derives from a potential; the second leads to smoothing of interfaces through surface tension (see e.g.~\cite{BrayRev}); and $\eta$ is a Gaussian white noise with zero mean and unit variance. A key feature of this equation is that all terms vanish in (open) regions where $\phi \equiv \pm 1$. Such regions are \emph{locally} absorbing: changes in $\phi$ may then only take place at domain boundaries.

The static phase behavior is found to depend on the shape of the potential, i.e., the values of $a$ and $b$ in (\ref{HammalLE}) \cite{Hammal2005}.  Here we are interested in what happens when the potential is held fixed, but the noise strength $\sigma$ is varied.  As an alternative to direct numerical integration of the Langevin equation (\ref{HammalLE}), the method used in \cite{Hammal2005}, we construct a stochastic microscopic model that is described by this Langevin equation at the mesoscopic scale. 

\begin{figure}[tb]
  \includegraphics[scale=0.85]{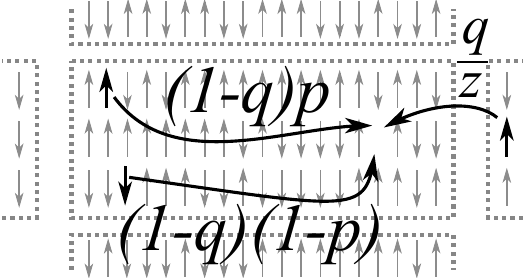}
\caption{Dynamics of the microscopic model. Each lattice site (shown as a region bounded by a dashed line) contains $N$ spins. In each update, a randomly-chosen spin is replaced with a copy taken from one of the $z=4$ nearest-neighbor sites, each with probability $q/z$; an up-spin from the same site with probability $(1-q)p$ or a down-spin from the same site with probability $(1-q)(1-p)$.}
\label{dynamics}
\end{figure}

We first define the dynamics of the microscopic model, and then show that it is described by a spatially-discrete version of (\ref{HammalLE}). It is defined on an $L\times L$ square lattice, each site of which hosts $N$ binary spins.  In each update, a spin is chosen at random and replaced with a copy from its neighborhood. To realize the diffusion term in (\ref{HammalLE}), there is a probability $q = \frac{hz}{N}$ that the copy is taken from one of the $z=4$ neighboring sites. Otherwise, the copy is taken from the same site. The potential term in (\ref{HammalLE}) can be interpreted as a \emph{systematic bias} towards copying one of the two spin states when copying from the same site. Specifically, if $b=0$, one should choose an up-spin with probability $p=\frac{1}{2}(1+\phi_i)+ \frac{2a}{N} \phi_{i}(1-\phi_i^2)$, where $\phi_i$ is the local magnetization. That is, $\phi_i = (n_{\uparrow, i} - n_{\downarrow, i})/N$, where $n_{\uparrow,i}$ and $n_{\downarrow,i}$ are the numbers of up and down spins on site $i$.  These update rules are illustrated in Fig.~\ref{dynamics}.

One concrete application of this model is in linguistics \cite{Baxter2009}. Each site represents a speaker, and each spin represents a previously-heard utterance stored in memory. The two spin states relate to two different ways of saying the same thing (e.g., phonetic realizations of a vowel). The local bias, in which the majority spin state is favored, then models an experimentally-observed tendency for language users to overproduce the most frequent variant \cite{hudsonkam}.  The key point is that the number of spins per site, $N$, sets both the noise strength (see below) \emph{and} the lifetime of an utterance in memory.  An appropriate choice for this lifetime was a central consideration in an analysis of the New Zealand English dialect \cite{Baxter2009}.

The stochastic equation of motion for the system is obtained by calculating moments of the change $\delta \phi_i$ in the local magnetization after one update \cite{Risken,Usm}.  All of these \emph{jump moments} vanish in the limit $N\to\infty$ as some power of $1/N$.  The limits $\lim_{N\to\infty} N^2 \langle \delta \phi_i \rangle = {\cal A}_i(\{\phi\})$ and $\lim_{N\to\infty} N^2 \langle \delta \phi_i^2 \rangle = {\cal B}_i(\{\phi\})$ are finite, in which
\begin{equation}
{\cal A}_i = a \phi_i (1-\phi_i^2) + h \sum_j \left(\phi_j - \phi_i\right)  \;,\quad {\cal B}_i = 2(1 -\phi_i^2) \;,
\end{equation}
and the sum is over nearest neighbors of site $i$. All other jump moments vanish at least as fast as $1/N^3$.  A Fokker-Planck equation for the model is then obtained from a Kramers-Moyal expansion \cite{Risken}. We define a time step as $\delta t\equiv 1/N^{2}$ and take the limit $N\rightarrow \infty$ to obtain a continuous time limit.  The Fokker-Planck equation
\begin{equation}
\label{FPE}
\partial_t P(\{\phi\}, t) = - \sum_i \partial_{\phi_i} [{\cal A}_i P]  + \frac{1}{2} \sum_i \partial_{\phi_i}^2 [{\cal B}_i P]
\end{equation}
exactly describes the stochastic dynamics of our microscopic model in this continuous-time limit.  When running stochastic simulations of the microscopic dynamics we set $N=100$ to allow for practicable run times.

This Fokker-Planck equation can be written equivalently (and thus also exactly) as a Langevin equation under the It\^o prescription \cite{Risken}. We have
\begin{equation}
 \partial_{t}\phi_{i}=h\bigg(r\phi_{i}(1-\phi_{i}^{2})+\sum_{j}(\phi_{j}-\phi_{i})\bigg)+\sqrt{1-\phi_{i}^{2}}\hspace{1mm} \eta \label{multLE} 
\end{equation}
which is a \emph{spatially-discrete} version of Eq.~(\ref{HammalLE}) with $a=rh$, $b=0$, $D = h\delta^2$, where $\delta$ is the lattice spacing, and $\sigma=1$. Written in this way it is clear that if $h$ is varied while holding $r$ constant then the strength of the noise, relative to the deterministic term, can be varied without changing the shape of the potential. We thus study the effect of noise strength in our Monte Carlo simulations of the microscopic dynamics by varying $h$. In all these simulations we take $r=3$, since we have found this to be large enough to provide the metastable interfaces between regions of positive and negative $\phi$ that are required if domain coarsening is to take place \footnote{We note a rich variety of behavior in the deterministic system at smaller $r$ that we do not discuss further here.}.

We now examine the dynamics in the high (small $h$) and low (large $h$) noise regimes. On a finite system, and with $a>0$, $b=0$ in (\ref{HammalLE}), the system eventually ends up in one of the globally absorbing states (i.e., all $\phi_i=1$ or all $\phi_i=-1$) \cite{Hammal2005}.  In the large-$h$ regime, we anticipate a domain growth driven by surface tension. This is because the deterministic limit $h\to\infty$ corresponds to the time-dependent Ginzburg-Landau equation, obtained from the Landau free energy functional for the Ising model with a non-conserved order parameter \cite{BrayRev}. When the noise amplitude is small, we do not expect the distinction between additive and multiplicative noise to be important, and thus the dynamics should coincide with model A for the Ising model.  On the other hand, it can be shown that in the limit $h\to0$, the purely fluctuation-driven dynamics of the voter model are formally recovered \cite{Mohle}. This suggests the possibility of at least a crossover (if not a transition) from Ising to voter coarsening as $h$ is reduced.

\begin{figure}[tb]
  \includegraphics[width=0.49\linewidth]{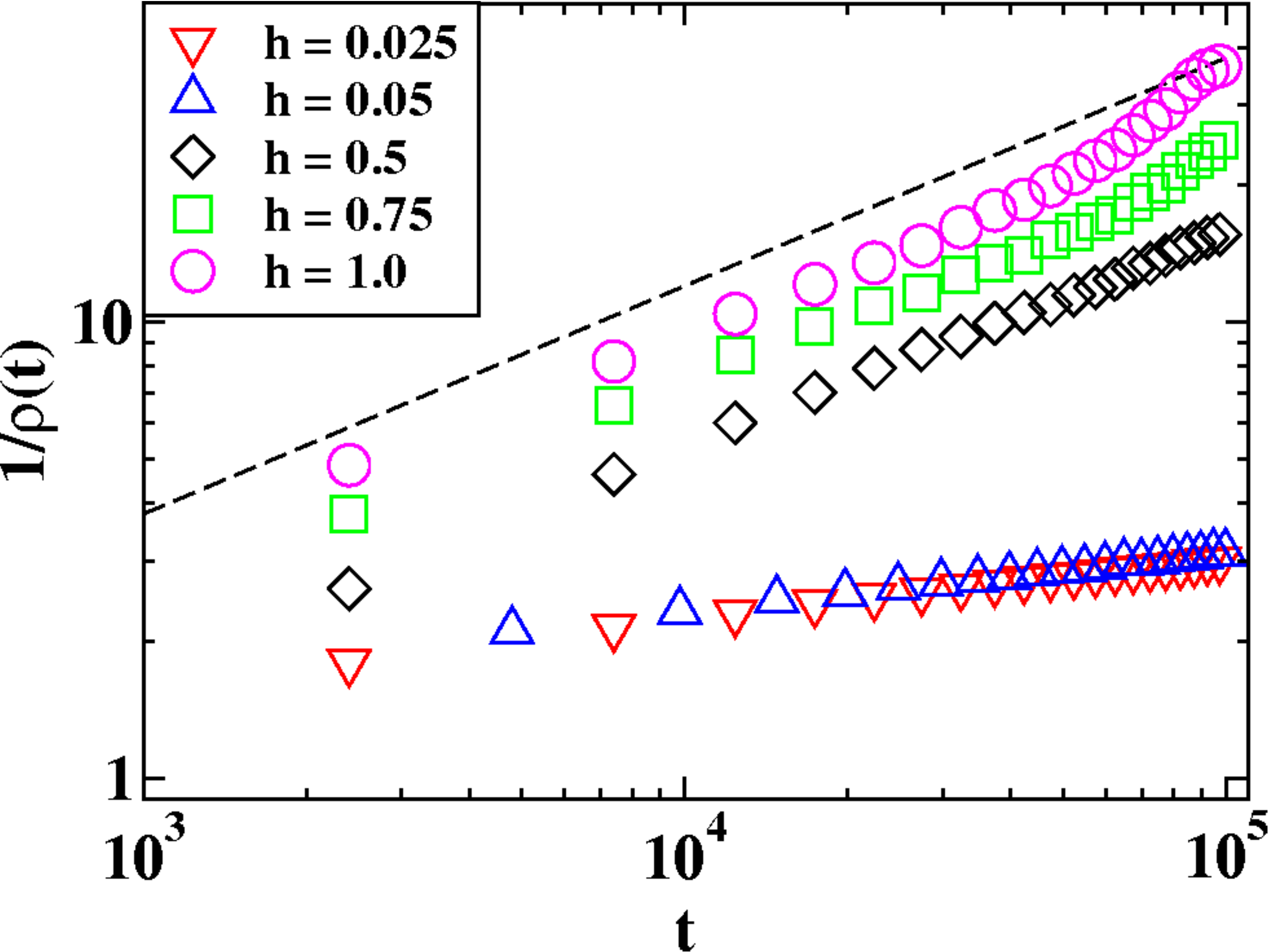}\hfill
  \includegraphics[width=0.49\linewidth]{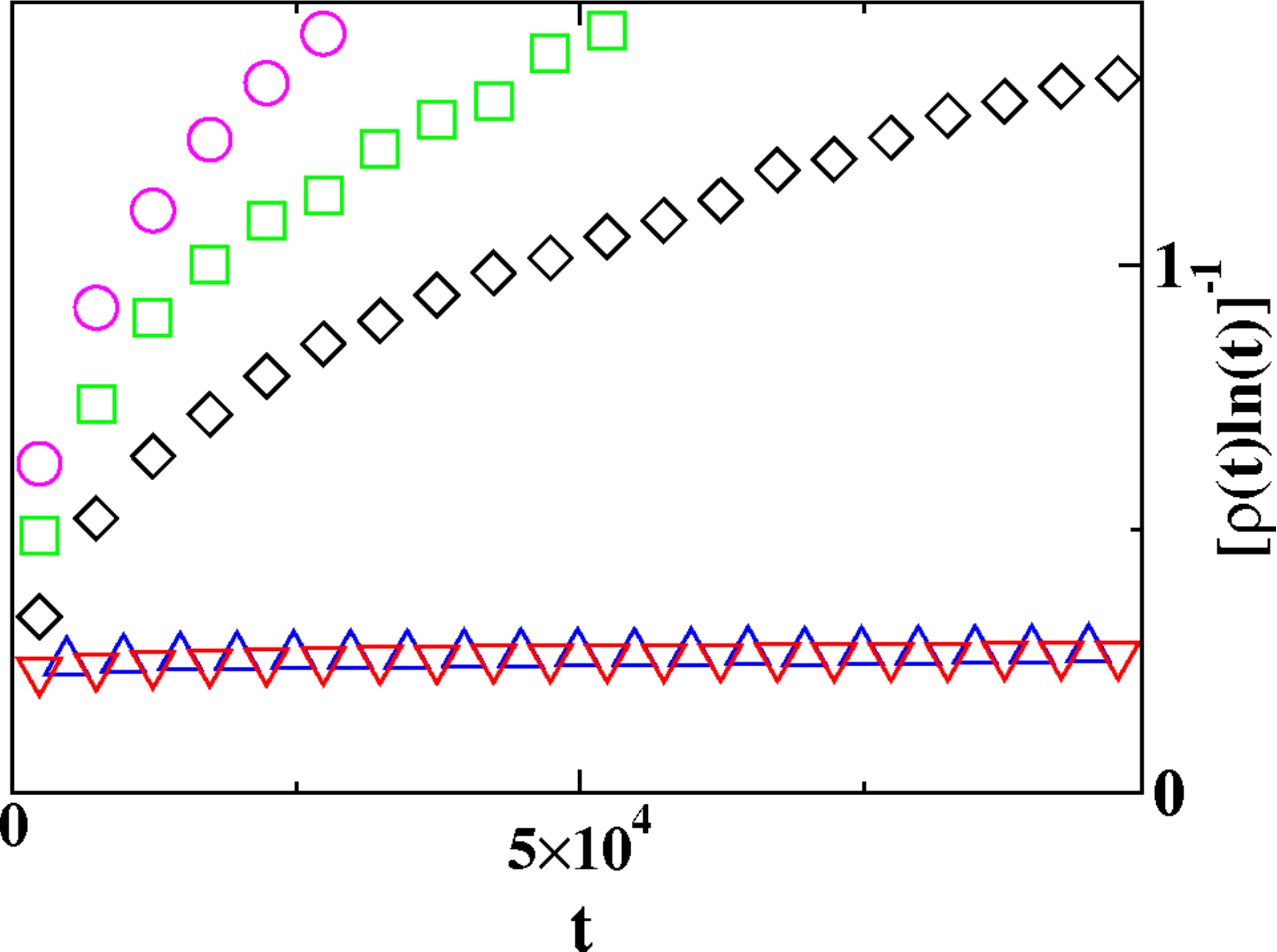}
\caption{(Color online) Interfacial density $\rho$ as a function of time $t$ in Monte Carlo simulations of the microscopic model. Left: $1/\rho(t)$ is plotted on logarithmic axes, such that a straight line indicates algebraic coarsening. For large $h$, and at times before the onset of finite-size effects, the gradient is consistent with the Ising model value of $\frac{1}{2}$ (dashed line).  Right: $[\rho(t) \ln t]^{-1}$ plotted on linear axes. A constant asymptote, seen for small $h$, indicates logarithmic coarsening.}
\label{densityplots}
\end{figure}

One way to identify the coarsening regime is to examine the density of interfaces as a function of time \cite{Dornic2001,Hammal2005}. This is defined as $\rho(t) = \frac{1}{4L^2} \sum_{\langle i,j \rangle} (1 - \phi_i \phi_j)$ on the square lattice in two dimensions, in which the sum is over distinct nearest-neighbor pairs. In the Ising model, $\rho(t) \sim t^{-1/2}$ \cite{BrayRev}, while in the voter model, $\rho(t)\sim 1/\ln(t)$ in two dimensions \cite{Krap1996}.  We measure $\rho(t)$ by averaging over multiple realizations of the microscopic stochastic dynamics described above. The results, shown in Fig.~\ref{densityplots}, are suggestive of algebraic coarsening taking place at large $h$, and logarithmic coarsening at low $h$. We remark that the deviation from the expected $t^{-1/2}$ law seen at late times for large $h$ is a finite-size effect caused by a domain coalescing with a periodic image of itself.

We now examine whether this shift from Ising to voter coarsening is a sharp transition or a smooth crossover, and if the former, what the mechanism for a transition would be.  The multiplicative noise in the Langevin equation (\ref{multLE}) makes interpretation of the stochastic dynamics difficult as it stands. Our physical intuition is strongest when the noise is additive, and the dynamics can be viewed in terms of diffusion in a potential. To this end, we transform the local magnetization $\phi_i$ to a variable $\theta_i$ such that the Fokker-Planck equation analogous to (\ref{FPE}) has a diffusion term ${\cal B}_i(\{\theta\})$ that is independent of the coordinates $\theta_i$.  The appropriate transformation is $\theta_i = \sin\phi_i$ (see~\cite{BlytheApp} for details). The resulting Fokker-Planck equation corresponds uniquely to the set of Langevin equations $\dot{\theta}_i = - V_{\rm D}'(\theta_i) + \eta$, one for each site $i$, and where $\eta$ is the usual thermal (Gaussian white) noise. The potential $V_{\rm D}$ has the form
\begin{equation}
V_{\rm D}(\theta)=\ln\left[\left(\frac{1-\sin \theta}{1+\sin \theta}\right)^{\frac{1}{2}hz m}(\cos \theta)^{\frac{1}{2}-hz}\right]+\frac{hr}{8}\cos2\theta \label{diffpot}
\end{equation}
where $m$ is the mean magnetization of the $z$ sites that are neighbors of site $i$. Note the $L^2$ separate Langevin equations are coupled because $m$ depends on the values of $\theta$ at neighboring sites.

The subscript ${\rm D}$ here is used to emphasize the crucial difference between this potential, felt by a thermal diffusion process, and that whose derivative gives the deterministic term in (\ref{multLE}).  The key points are that the shape of $V_{\rm D}$ gives insight into the stability of a microscopic configuration under the stochastic dynamics, and that this shape changes with the noise strength $h$ in a nontrivial way.  For our purposes it suffices to look at the behavior near the boundaries, from which we deduce three distinct qualitative shapes that this \emph{diffusion potential} may take.

\begin{figure}[tb]
 \subfigure[\hspace{1mm}$h<h_{-}$]{\label{Potentialc}\includegraphics[width=0.32\linewidth]{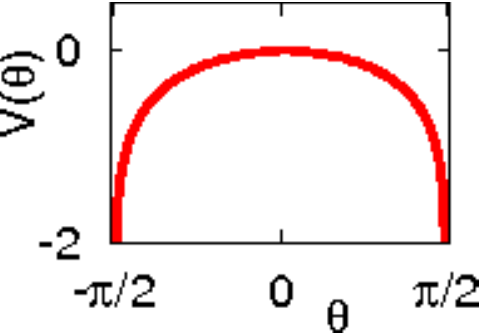}}\hfill
  \subfigure[\hspace{1mm}$h_{-}<h<h_{+}$]{\label{Potentialb}\includegraphics[width=0.32\linewidth]{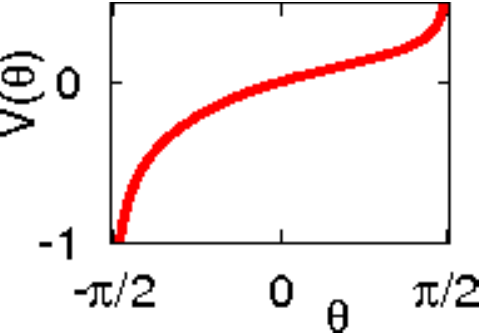}}\hfill
  \subfigure[\hspace{1mm}$h>h_{+}$]{\label{Potentiala}\includegraphics[width=0.32\linewidth]{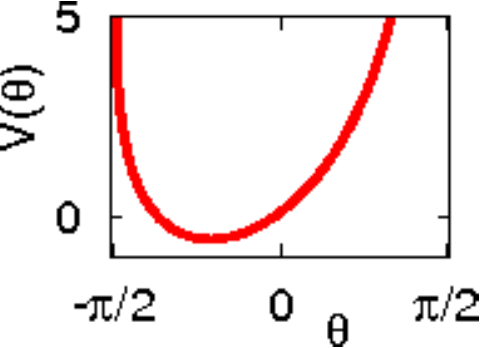}}
\caption{(Color online) Shape of the diffusion potential at increasing $h$.  Here, $h_{\pm}=[2z(1\pm\bar{m})]^{-1}$, $z=4$ and $\bar m=-0.5$.}
\label{newpots}
\end{figure}

As $\theta$ approaches either boundary point $\theta = \pm \frac{\pi}{2}$ at some fixed local magnetization $\bar{m}$, $V_{\rm D}$ diverges logarithmically. Depending on the values of $h$ and $\bar{m}$, the divergence can be either towards $+\infty$ or $-\infty$.  Considering the case $\bar{m}<0$, we find that the divergence is towards $+\infty$ at the right boundary when $h > h_{-} = [2z(1-\bar{m})]^{-1}$, and towards $-\infty$ for smaller values of $h$.  Similarly, at the left boundary, the divergence is towards  $+\infty$ when $h > h_{+} = [2z(1+\bar{m})]^{-1}$.  The three possible combinations of boundary divergence are as shown in Fig.~\ref{newpots}. Analogous shapes are found for the case $\bar{m}>0$ by using the symmetry $\theta_i \to -\theta_i$ and $\bar{m} \to -\bar{m}$.

We contend that this changing shape of the diffusion potential is of significance for the coarsening dynamics for the following reason.  For any microscopic spin configuration, we always have that $h_{-} > \frac{1}{4z}$. Therefore, for $h<\frac{1}{4z}$, there are always minima of the diffusion potential at the boundaries, and a single maximum in between (since there are at most two extrema in the interior region). Thus the diffusion of $\theta$ is at any instant biased towards a locally absorbing state. Once a locally absorbing state is reached, the deterministic force in Eq.~(\ref{multLE}) vanishes. Then, all that remains is the Langevin equation for the voter model (the form of which is given e.g.~in \cite{Hammal2005}).  This leads us to suggest that voter coarsening should be seen over some \emph{finite} range of $h$, at least up to $h=\frac{1}{4z}$.  Since the deterministic limit of (\ref{multLE}) and model A for the Ising model are equivalent, there is the possibility of a transition to Ising-type coarsening at some nonzero value of $h$, perhaps in the vicinity of $h=\frac{1}{4z}$.

This picture is confirmed if we can identify a point at which the ordering dynamics changes from the voter-type to the Ising-type from Monte Carlo simulation data. Here we have found a \emph{droplet experiment}, similar to that described in \cite{Dornic2001}, most useful. This entails an initial condition in which all spins within a radius $R_0$ of the origin are up, and the rest are down.  When coarsening under surface tension, the Allen-Cahn theory of phase ordering predicts that the droplet will shrink, its area decreasing at a constant rate proportional to the diffusion constant $D$ \cite{BrayRev}.  Since $D \propto h$ in our model, we expect the magnetization density $m(t) = \frac{1}{L^2} \sum_{i} \phi_i$ to decrease at a rate $c \propto h/L^2$ through surface tension. On the other hand, in the voter model, the rate of change of the magnetization, averaged over multiple stochastic realizations, is zero.

\begin{figure}[t]
\includegraphics[width=0.8\linewidth]{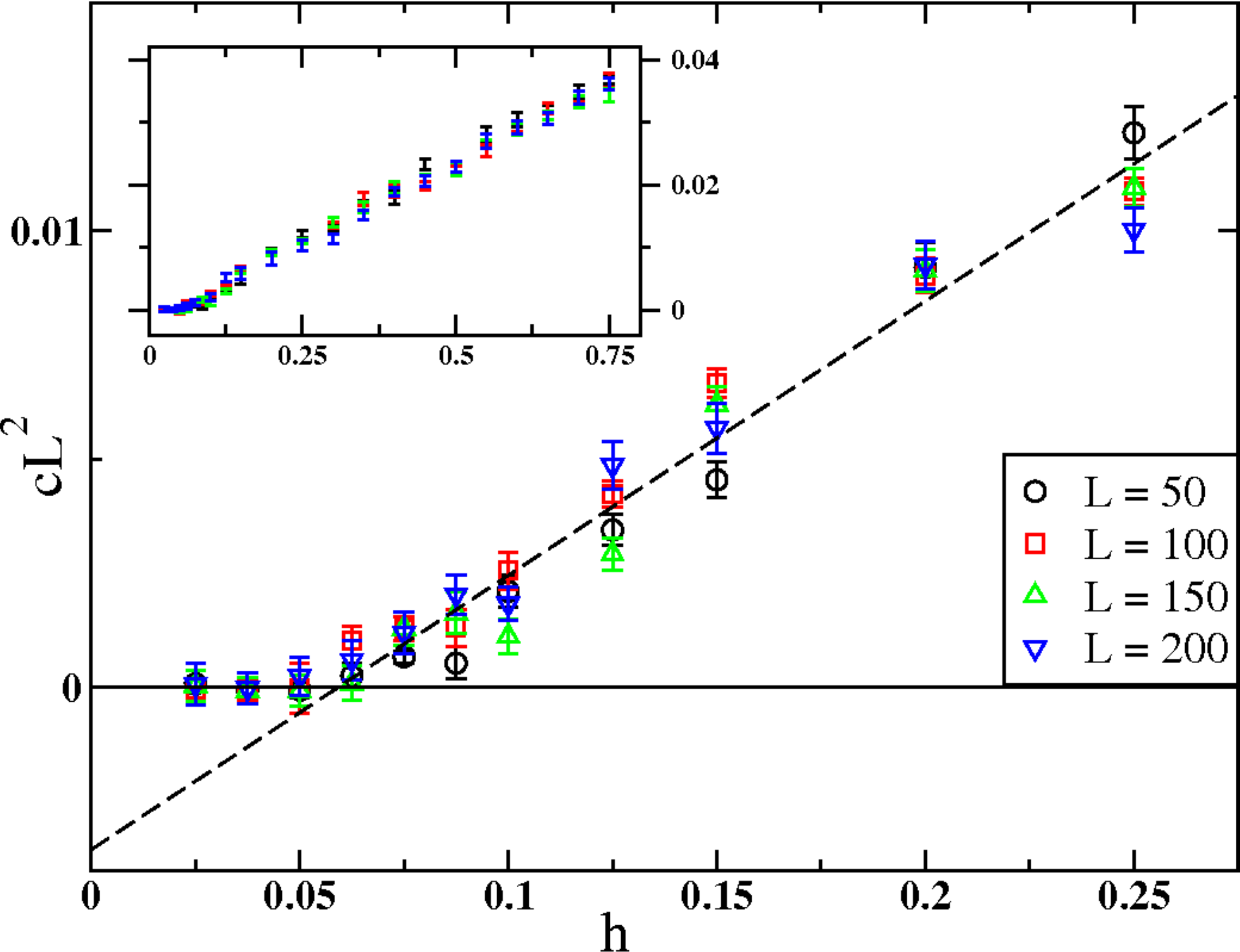}
\caption{(Color online) Droplet shrinking rate, $c$, as a function of inverse noise strength $h$ at a range of system sizes $L$. Each data point is obtained from a sample of 100 simulation runs.}
\label{magplot}
\end{figure}

To each stochastic realization of $m(t)$ we fit a linear function. An estimate of $c$, and an error, can be computed from the mean and standard deviation of the measured gradients.  We can compare results from different system sizes $L$ by taking the initial droplet radius $R_0$ proportional to $L$, and by plotting $cL^2$ against $h$.  These results are shown in Fig.~\ref{magplot}. The expected linear increase of $c$ with $h$ is observed. What is interesting is that the intercept is not at $h=0$, but at some positive value of $h$, below which $c$ is consistent with zero.  This suggests that there is indeed a transition between voter and Ising coarsening dynamics at some $h=h^*>0$.  A least-squares linear fit to the data yields an estimate of $h^* \approx 0.059$, which is close to the value $\frac{1}{4z} = 0.0625$ suggested by the analysis of the diffusion potential. It is possible that a nonzero transition point could be a finite-size effect. The fact that the data show no systematic shift of $h^*$ towards zero as $L$ is increased appears to rule this out.

To summarize, we have presented evidence for a noise-induced transition in the ordering dynamics in systems with two symmetric absorbing states and discrete space.  The origin of this transition is that, once the noise is sufficiently strong, \emph{locally}-absorbing states are entered with ease. This eliminates the term in (\ref{multLE}) that leads to surface-tension driven coarsening. Then, fluctuation-driven voter-type coarsening is the only possible way the system can order. We believe discrete space is essential for the operation of this mechanism, so that a locally-absorbing state ($\phi_i=\pm1$) can be reached more quickly than the diffusion process connecting neighboring sites. By contrast, in continuous space, any finite transition point $h^*>0$ collapses onto a zero diffusion constant, $D=0$. Therefore, one would generically expect Ising-type ordering on the continuum. This could explain why a previous renormalization group treatment of (\ref{HammalLE}) did not reveal the noise strength as a relevant quantity \cite{GallaVM}.

However, the fact the combination of discrete space and non-thermal noise leads to a further phase transition is in itself interesting. It would be worthwhile to try and understand this transition more rigorously than through the heuristic diffusion potential picture we have described here.  The fact that noise strength is related to memory lifetime in the application of our microscopic model to language change also warrants further investigation, in particular, whether the transition is also seen on social network structures and whether the resulting patterns of diversity can be distinguished empirically. 

We conclude by returning to a question investigated by \cite{Dornic2001}: that is, which \emph{qualitative} features, such as symmetries and conservation laws, determine the dynamical universality class of systems with multiple absorbing states.  We have seen that a set of models with the same qualitative properties, but different noise strengths, can have different ordering dynamics.  In particular, strong noise leads to the magnetization conservation characteristic of the voter model being an \emph{emergent consequence} of the stochastic dynamics.  To develop a better general understanding of nonequilibrium phase transitions and critical phenomena, one may need to determine whether other cases exist in which such conservation laws emerge.

This work has made use of the resources provided by the Edinburgh Compute and Data Facility (ECDF). We thank Ivan Dornic, Martin Evans, G\'{e}za \'{O}dor and Julien Tailleur for comments on the manuscript, and the EPSRC (D.I.R.) and RCUK (R.A.B.) for financial support.

\bibliography{Year1biblo}
\end{document}